\documentclass[10pt,journal]{IEEEtran}

\usepackage{mathtools, cases}
\usepackage{textcomp}
\usepackage{makecell}
\usepackage{ragged2e}
\usepackage{amsthm}
\usepackage{amsfonts,amssymb}
\usepackage{cite, url}
\usepackage{verbatim}
\usepackage{bm}
\usepackage{graphicx}
\usepackage{xcolor}
\usepackage{subcaption}
\usepackage{epstopdf}
\usepackage{mathrsfs}
\usepackage{txfonts}
\usepackage{lineno}
\usepackage{multirow}
\usepackage{booktabs}
\usepackage{color}
\usepackage{multicol}
\usepackage{stfloats}
\usepackage{diagbox}
\usepackage{esint}
\usepackage{float}
\usepackage{algpseudocode}
\usepackage{empheq}
\usepackage{tabularx}
\usepackage{enumitem}

\allowdisplaybreaks[4]

\begin{document}

\title{\color{black}Generative Semantic Communication via Textual Prompts: Latency Performance Tradeoffs}

\author{Mengmeng Ren, Li Qiao, Long Yang, Zhen Gao, Jian Chen, \\ Mahdi Boloursaz Mashhadi, Pei Xiao, Rahim Tafazolli, and Mehdi Bennis 
 \vspace{-10mm}
\thanks{Copyright (c) 20xx IEEE. Personal use of this material is permitted. However, permission to use this material for any other purposes must be obtained from the IEEE by sending a request to pubs-permissions@ieee.org. This work was supported in part by the National Natural Science Foundation of China under Grant 62371367, in part by the Key Research and Development Program of Shaanxi under Grant 2023-ZDLGY-50, in part by the Innovation Capability Support Program of Shaanxi under Grant 2024ZC-KJXX-080 and Grant 2024RS-CXTD-01, and in part by the U.K. Engineering and Physical Sciences  Research  Council  under Grant EP/X013162/1. The work of Mengmeng Ren was supported by China Scholarship Council. \emph{(Corresponding
author: Long Yang.)}}
\thanks{M. Ren, L. Yang, and J. Chen are with the State Key Laboratory of Integrated Services Networks, Xidian University, Xi'an 710071, China (e-mail: renmengmeng@stu.xidian.edu.cn; lyang@xidian.edu.cn; jianchen@mail.xidian.edu.cn).}
\thanks{L. Qiao is with the School of Information and Electronics, Beijing Institute of Technology, Beijing
100081, China (e-mail: qiaoli@bit.edu.cn). Zhen Gao is with the State Key Laboratory of Integrated Services Networks (Xidian University), Xi'an 710071, China, and is also with Beijing Institute of Technology, Beijing, 100081, China (e-mail: gaozhen16@bit.edu.cn).}
\thanks{M. Boloursaz Mashhadi, P. Xiao, and R. Tafazolli are with 5GIC \& 6GIC, Institute for Communication Systems (ICS), University of Surrey, GU2 7XH Guildford, U.K. (emails: \{m.boloursazmashhadi, p.xiao, r.tafazolli\}@surrey.ac.uk).  }
\thanks{M. Bennis
is with the Centre for Wireless Communications, University of Oulu, 90014
Oulu, Finland (e-mail: mehdi.bennis@oulu.fi).}
}

\maketitle

\begin{abstract}
This paper develops an edge-device collaborative Generative Semantic Communications (Gen SemCom) framework leveraging pre-trained Multi-modal/Vision Language Models (M/VLMs) for ultra-low-rate semantic communication via textual prompts. The proposed framework optimizes the use of M/VLMs on the wireless edge/device to generate high-fidelity textual prompts through visual captioning/question answering, which are then transmitted over a wireless channel for SemCom. Specifically, we develop a multi-user Gen SemCom framework using pre-trained M/VLMs, and formulate a joint optimization problem of prompt generation offloading, communication and computation resource allocation to minimize the latency and maximize the resulting semantic quality. Due to the non-convex nature of the problem with highly coupled discrete and continuous variables, we decompose it as a two-level problem and propose a low-complexity swap/leaving/joining (SLJ)-based matching algorithm. Simulation results demonstrate significant performance improvements over the conventional semantic-unaware/non-collaborative generation offloading benchmarks.
\end{abstract}
\vspace{-2mm}


\section{Introduction}
The recent integration of \textit{Large Generative Artificial Intelligence Models} with wireless networks provides ample opportunities to develop innovative technologies with transformative potential \cite{TelecomGen, NetGPT, LLM}. One such technology is \textit{Generative Semantic Communications (Gen SemCom)}, which leverages the capabilities of generative AI models to develop ultra-low bitrate semantic communication systems aiming to transmit only the semantic message of interest with high fidelity \cite{Liang2024generative, Wanting2024generative,10343094}. Semantic communications through \textit{textual descriptions} or \textit{prompts} has recently emerged as an efficient Gen SemCom scheme to convey the most important semantics of the source signal to the receiver in a compressed format as compact as a textual prompt, thereby achieving ultra-low-rate semantic communication \cite{qiao2024latency, Zhijin, cicchetti2024language}.

In this setup, pre-trained \textit{Multi-modal/Vision Language Models (M/VLMs)} are used to generate a high-fidelity textual prompt from the source signal, i.e. visual captioning/question answering, that preserves its intended semantic content. This textual prompt is transmitted to the receiver, making its semantic quality critical. A larger M/VLM can generate a higher fidelity prompt that can be better adjusted to the communication intent, at the cost of significantly increased computation complexity and latency. On the other hand, smaller versions of these  M/VLMs that can run \textit{on-device},  will reduce the computation complexity/latency, but compromise the quality of the generated prompt. Moreover, considering the computing resources available on the edge and device, the computations required for prompt generation can be carried out either with smaller models on-device, or offloaded to the edge servers where larger M/VLMs are deployed.
{Recently, the semantic-unaware computation offloading mainly focuses on optimizing communication and computation resources\cite{2024TVT,2024JSAC1,2024JSAC2}. Specifically, the work in \cite{2024TVT} proposed a block coordinate descent-based algorithm to minimize user latency. Furthermore, to minimize the energy consumption, a long-term queue-aware scheme was proposed in \cite{2024JSAC1}, and an iterative algorithm based on successive convex approximation was proposed in \cite{2024JSAC2}, respectively. Unlike existing semantic-unaware studies\cite{2024TVT,2024JSAC1,2024JSAC2} that solely focus on minimizing the latency or energy consumption, Gen SemCom design should also incorporate semantic-related performance metrics, which motivates our work.}

In this paper, {we develop a \textit{multi-user edge-device collaborative Gen SemCom framework} via textual prompts, leveraging the computation capabilities and diverse pre-trained M/VLMs available on edges and devices.} Unlike existing SemCom studies \cite{SemCom1, Semcom2} that employ end-to-end training of the transmitter/receiver, our framework adopts \textit{pre-training and generation offloading} to avoid the computationally intensive end-to-end training on
edge/device, utilizing the \textit{zero/few-shot performance} \cite{FSL1} of the M/VLMs pre-trained on large data corpora. The main contributions are summarized as follows.
\begin{itemize}
    \item For the proposed framework, we formulate a joint optimization of prompt generation offloading and communication/computation resource allocation to reduce the total SemCom latency while maximizing the corresponding semantic quality. The formulated problem belongs to mixed integer nonlinear programming (MINLP) problems that are mathematically intractable. To solve this problem, we first equivalently decompose it into a two-level problem that iteratively solves the discrete and continuous variables.
    \item For the formulated two-level problem, we propose a low-complexity swap/leaving/joining (SLJ)-based matching algorithm. Unlike conventional matching algorithms\cite{matching} that are sensitive to initial matchings, our proposed algorithm reduces the dependency on initial matchings by the designed leaving and joining operations, which can significantly enhance the performance.
    \item Simulation results demonstrate the superiority of the proposed framework over semantic-unaware/non-collaborative generation offloading benchmarks in terms of the total SemCom latency and quality, achieving an average performance enhancement of at least 21.73\%.
\end{itemize}
\vspace{-1em}
\section{Edge-Device Collaborative Gen SemCom via M/VLMs}
\subsection{M/VLMs Enabled Edge-Device Collaborative Gen SemCom}
Various pre-trained M/VLMs have significant capabilities for zero/few-shot generation of the textual prompt with the limited locally provided context in a visual captioning or visual question answering setup. These models are typically pre-trained on extensive datasets containing image-text pairs, and thereby can understand and generate language outputs grounded in visual inputs. Driven by the semantic extraction capability provided by these M/VLMs, we consider a multi-user Gen SemCom system consisting of $K$ edge servers and $N$ semantic transmitter-receiver (T-R) pairs. {\color{black}For each T-R pair, the aim is to transmit a textual prompt to the receiver that contains the most significant semantic contents of the source signal (e.g., image, video, point cloud, etc) at the transmitter\footnote{{Depending on the application scenario, the textual prompt may either be directly used, e.g. in human machine interaction (HMI) applications like human-supervised control of robotic tele-operations \cite{AutoRT}, or be input to a generative text-to-image model to locally synthesize a semantically consistent signal at the receiver for ultra-low-rate semantic communications, e.g. extended/mixed reality (XR/MR) \cite{XR}, holographic teleportation, immersive communications and the internet of senses \cite{IoS}.}}.} Let $A_k$ ($k \in \mathcal{K} \triangleq \{ 1, \ldots, K\}$) denote the $k$-th edge server, and $\{t_n, r_n\}$ ($ n \in \mathcal{N}\triangleq \{1, \ldots, N\}$) denote the $n$-th semantic T-R pair, where $t_n$ and $r_n$ denote the transmitter and the receiver, respectively. {Inspired by \cite{integ}, the integration of M/VLMs into this wireless system is designed as the following three stages.}

{1) \emph{Initial network preparation stage}: First, M/VLMs (e.g., CapPa, Flamingo, BLIP, etc.) are pre-trained on extensive datasets containing image-text pairs. The architecture\footnote{Note that the performance of M/VLMs is evaluated using a variety of metrics, such as bilingual evaluation understudy (BLEU), consensus-based image description evaluation (CIDEr), etc.} of these M/VLMs for prompt generation follows an encoder-decoder structure, e.g., a Vision Transformer (ViT) encoder is combined with a Transformer-based text decoder, where different cross/self-attention mechanisms are utilized for inter/intra-modal features.
After pre-training, different M/VLMs are downloaded and deployed on devices and edge servers. {{Specifically, each transmitter is equipped with a lightweight M/VLM acting as a semantic encoder, denoted by $F_{E,n}(\cdot|\theta_n^\ast)$, each server is equipped with a large-scale M/VLM serving as a semantic encoder, denoted by $F^{\prime}_{E,k}(\cdot|w_k^\ast)$, where $\theta_n^\ast$ and $w_k^\ast$ are parameters of pre-trained M/VLMs. Moreover, each receiver is equipped with a generative decoder, e.g., stable diffusion \cite{qiao2024latency}, denoted by $F_{D,n}(\cdot|\xi_n^\ast)$, where $\xi_n^\ast$ is the pre-trained parameters. }}}

{
2) \emph{M/VLMs-integrated SemCom service provisioning stage}: Once M/VLMs are successfully deployed, Gen SemCom transmission is conducted based on network optimization results. For on-device prompt generation, transmitter $t_n$ locally extracts semantic prompt $s_n$ using model $F_{E,n}$, given by $s_n = F_{E,n}(\bar{X}_n|\theta_n^\ast)$, where $\bar{X}_n$ denotes the source signal. The extracted prompt $s_n$ is then encoded into bitstream $X^{\prime}_n$ using a text encoding method $\bar{B}_t$, e.g., UTF-8 \cite{yergeau2003utf}, represented by $X^{\prime}_n = \bar{B}_t(s_n)$. Subsequently, $X^{\prime}_n$ is modulated and transmitted over wireless channel.
For offloaded prompt generation, transmitter $t_n$ directly encodes $\bar{X}_n$ into bitstream $X_n$, given by $X_n = \bar{B}_i(\bar{X}_n)$, where $\bar{B}_i$ denotes the encoding method, e.g., JPEG algorithm\cite{wallace1991jpeg} for image encoding. Then, $X_n$ is modulated and transmitted to selected server $A_k$ over wireless channel. Next, server $A_k$ decodes received signal ${\tilde{X}}_{n}$ and extracts its semantic prompt using $s^{\prime}_{n,k}=F^{\prime}_{E,k}(\bar{B}^{-1}_i(\tilde{X}_n)|w_k^\ast)$, where $\bar{B}^{-1}_i$ is the inverse operation of $\bar{B}_i$. The extracted prompt $s^{\prime}_{n,k}$ is then encoded into bitstream $X^{\prime}_{n,k} = \bar{B}_t(s^{\prime}_{n,k})$, and is modulated and transmitted to the receiver.
Finally, at receiver $r_n$, the received bitstream, i.e., $\hat{X}^{\prime}_n$ for notation convenience, is converted into prompt using $\hat{s}_n =\bar{B}^{-1}_t(\hat{X}^{\prime}_n)$, where $\bar{B}^{-1}_t$ is the inverse operation of $\bar{B}_t$. Afterwards, the receiver can directly use the prompt for semantic-based control tasks or regenerate the original signal using decoder $F_{D,n}$, i.e., $\hat{X}_n = F_{D,n}(\hat{s}_n |\xi_n^\ast)$.
}

{
3) \emph{Model update stage}: Following the similar procedure as in the \emph{initial network preparation stage}, transmitters, receivers, and edge servers will periodically update M/VLMs parameters and remove/load M/VLMs according to the requirements of network operators.}
\vspace{-1mm}
\begin{figure}[t]
		\centering  \includegraphics[width=0.4 \textwidth]{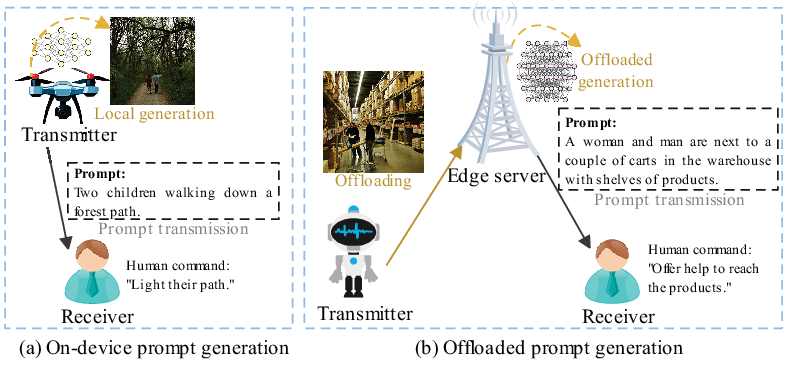}
  \captionsetup{font={footnotesize}, singlelinecheck = off, justification = justified, name={Fig.},labelsep=period}
  \caption{{Edge-device collaborative Gen SemCom via textual prompts in a typical application, i.e. human-supervised control of tele-operations: (a) on-device prompt generation; (b) offloaded prompt generation.}}
  \vspace{-7mm}
		\label{SystemDiag}
\end{figure}
\vspace{-1em}
\subsection{On-Device vs. Offloaded Prompt Generation}
In this framework, the computation model of lightweight M/VLM deployed at transmitter $t_n$ is characterized by $\mathcal{M}_n \triangleq \{F_{E,n}(\cdot|\theta_n^\ast), F_n, I_n, Q_n\}$, where $F_n$ denotes the required computations in floating point operations (FLOPs) of the model at $t_n$, $I_n$ (cycles$/$FLOP) denotes the computation intensity for conducting one floating point operation, and $Q_n$ denotes the corresponding semantic quality when the textual prompt is generated locally at $t_n$. Likewise, the computation model of large-scale M/VLM deployed for \textcolor{black}{transmitter $t_n$}'s prompt generation at edge server $A_k$ is characterized by $\mathcal{M}^\prime_k \triangleq \{F^{\prime}_{E,k}(\cdot|w_k^\ast), F^\prime_k, I^\prime_k, Q^\prime_{n,k}\}$, where $F^\prime_k$ denotes the required computations in FLOPs of the corresponding model at edge server $A_k$, $I^\prime_k$ (cycles$/$FLOP) denotes the corresponding computation intensity, and $Q^\prime_{n,k}$ denotes the semantic quality\footnote{{For the proposed framework, the role of semantic communications is not only assigning values to $Q_n$ and $Q^{\prime}_{n,k}$, but also leveraging high-quality pre-trained M/VLMs to extract semantic information from raw data, thereby enabling universal intent- and task-aware transmissions.}} when the prompt generation task for $t_n$ is offloaded to edge server $A_k$.

As depicted in Fig. \ref{SystemDiag}, each transmitter can choose to generate the prompt on-device using its locally deployed lightweight M/VLM, and transmit the prompt to the paired receiver. Alternatively, the transmitter may choose to transmit the compressed source signal to a selected edge server, which subsequently generates the prompt by the deployed large-scale M/VLM and forwards the prompt to the paired receiver. Specifically, the offloading decision of transmitters is expressed as $\alpha_n \in \{0, 1\}$ ($\forall n \in \mathcal{N}$), where $\alpha_n = 0$ indicates transmitter $t_n$ locally generates the prompt, otherwise, $\alpha_n = 1$. Moreover, if transmitters choose to offload prompt generation to an edge server, the association relationship between transmitters and edge servers is expressed by $\beta_{n, k}$ ($n \in \mathcal{N}, k \in \mathcal{K}$), where $\beta_{n,k} = 1$ indicates $t_n$ is served by edge server $A_k$, otherwise, $\beta_{n,k}=0$. We assume that all the transmitters/receivers/edge servers are equipped with a single antenna. All the wireless channels experience independent but non-identically distributed Rayleigh block-fading, indicating the channel gains remain unchanged within one transmission block but may change independently over different blocks. The channel gains of links $t_n \to r_n$, $t_n \to A_k$, and $A_k \to r_n$ are denoted by $|h_n|^2$, $|h_{n,k}^u|^2$, and $|h_{k,n}^d|^2$, respectively.

\vspace{-1em}
\subsection{Latency/Performance Tradeoffs}
If transmitter $t_n$ chooses to locally generate the prompt from source signal, the corresponding computation latency is given by $\tau_{n}^L = \frac{F_n I_n}{f_n^L}$, where $f_n^L$ denotes the device computation frequency (cycles/s) for $t_n$. Accordingly, the energy consumption for this local computation is $e_n^L = \kappa_n (f_n^L)^3 \tau_n^L = \kappa_n F_n I_n (f_n^L)^2$, where $\kappa_n$ is the effective capacitance coefficient related to the device processor of $t_n$ \cite{4DMatching}. After locally generating the prompt, transmitter $t_n$ will send it to its paired receiver with the transmission rate $R_n^{T-R} = B \log_2(1+\frac{p_n |h_n|^2}{\sigma_n^2})$, where $p_n$ is the transmit power of $t_n$, $\sigma_n^2$ is the additive white Gaussian noise (AWGN) power received at $r_n$, and $B$ is the bandwidth. Accordingly, the prompt communication latency is $\tau_n^{T-R} = \frac{|X_n^\prime|}{R_n^{T-R}}$, and the resulting communication energy consumption is $e_n^{T-R} = p_n \tau_n^{T-R}$, where $|X_n^\prime|$ is the data size (bits) of prompt $X_n^\prime$.

Alternatively, if transmitter $t_n$ chooses to offload the prompt generation task to a selected edge server, e.g. $A_k$, the transmission rate for $t_n$ to offload the compressed source signal to the server is $R_{n,k}^{u} = B \log_2(1+\frac{p_n |h_{n,k}^{u}|^2}{\sigma_{n,k}^2})$, and the corresponding communication latency is $\tau_{n,k}^u = \frac{|X_n|}{R_{n,k}^{u}}$, where $\sigma_{n,k}^2$ is the AWGN power received at edge server $A_k$, and the corresponding energy consumption is $e_{n,k}^u = p_n \tau_{n,k}^u$. After receiving the compressed source signal, the edge server $A_k$ will generate the prompt using its deployed large-scale M/VLM, and the corresponding computation latency is $\tau_{n,k}^e = \frac{F^{\prime}_k I^\prime_k}{f_{n,k}^e}$, where $f_{n,k}^e$ is the allocated computation frequency of $A_k$ to generate the textual prompt. Next, when $A_k$ generates the textual prompt, it will directly transmit it to the corresponding paired receiver $r_n$ with transmission rate $R_{k,n}^d = B \log_2(1+\frac{p_{k,n|h^d_{k,n}|^2}}{
\hat{\sigma}_{k,n}^2})$, where $p_{k,n}$ is the allocated transmit power of $A_k$ and $\hat{\sigma}_{k,n}^2$ denotes the AWGN power received at $r_n$. Accordingly, the communication latency for sending the prompt to $r_n$ is $\tau_{k,n}^d = \frac{|X_{n,k}^\prime|}{R_{k,n}^d}$, where $|X_{n,k}^\prime|$ is the data size of $X_{n,k}^\prime$. Since edge servers are generally constantly charged, we only consider the energy consumption minimization of transmitters.

Based on the above, the total semantic communication latency between the T-R pair is given by $\tau_n = (1-\alpha_n) (\tau_n^L + \tau_n^{T-R}) + \alpha_n \sum_{k=1}^K \beta_{n,k} (\tau_{n,k}^u + \tau_{n,k}^e +\tau_{k,n}^d  ) $, and the corresponding energy consumption is $e_n =  (1-\alpha_n)(e_n^L + e_{n}^{T-R}) + \alpha_n (\sum_{k=1}^K \beta_{n,k}e_{n,k}^u)$. Considering different zero/few-shot performance of the edge and device M/VLMs, the resulting semantic quality of the generated prompt is $(1-\alpha_n)Q_n + \alpha_n \sum_{k=1}^K \beta_{n,k}{Q_{n,k}^\prime}$. Then, we define a new metric, termed \emph{communication, computation and quality (CCQ)}, which jointly considers the SemCom latency and the semantic quality. {To mitigate the impact of different orders of metric magnitude on overall performance, the CCQ metric is defined as $m_n^{CCQ} \triangleq \frac{\omega_n \tau_n}{(1-\alpha_n)\hat{Q}_n + \alpha_n \sum_{k=1}^K \beta_{n,k}{\hat{Q}_{n,k}^\prime}}$, where $\omega_n$ denotes the weight factor of latency, $\hat{Q}_n = {(Q_n - Q_{\min})}/{(Q_{\max}- Q_{\min})}$ and $\hat{Q}_{n,k}^\prime= ({{Q}_{n,k}^\prime - Q_{\min}})/({Q_{\max}- Q_{\min}})$ denote the normalized semantic quality with $Q_{\min}$ and $Q_{\max}$ being predefined constants, respectively. The physical meaning of the CCQ metric is to jointly evaluate both the timeliness of the delivered information and the precision of the extracted semantic content. Unlike conventional semantic performance metrics (e.g., CIDEr) that focus solely on the precision/accuracy, our proposed metric  measures both timeliness and precision of the received information, which is well-suited for communication systems in dynamic environments, e.g., HMI, intelligent transportation, and surveillance systems.}%

\vspace{-1em}
\subsection{Semantic-Aware Generation Offloading Problem}
Considering fairness among the transmitters, we focus on minimization of the maximal CCQ among T-R pairs by jointly optimizing the M/VLM prompt generation offloading strategy, as well as the transmit powers and computation frequencies of the transmitters and edge servers. Accordingly, the semantic-aware optimization problem is formulated as
\begin{subequations}
\label{formulationProblem}
\vspace{-1pt}
\begin{align}
\label{eq_OP1} & \quad \min_{\substack{\alpha_n, \beta_{n,k}, p_n, f_n^L, p_{k,n}, f_{n,k}^e}}  \max_{\forall n} m_n^{CCQ} \\
\label{OP1_1}      &\quad  \text{s.t.}  \alpha_n \in \{0,1\}, \: \beta_{n,k} \in \{ 0, 1 \}, \forall n\in \mathcal{N}, \forall k \in \mathcal{K}, \\
\label{OP1_3}&   \quad \quad  \sum\nolimits_{k=1}^K \beta_{n,k}=1,   \forall n \in \mathcal{N},  \\
\label{OP1_4}&   \quad \quad  \sum\nolimits_{n=1}^N \alpha_n \beta_{n,k} \leq N_k^{\max}, \forall k \in \mathcal{K},
\\
\label{OP1_8}&   \quad \quad
0 \leq p_n \leq p_n^{\max}, \: 0 \leq f_n^L \leq f_{n, \max}^L, \forall n \in
\mathcal{N}, \\
\label{OP1_6}&   \quad \quad
\sum\nolimits_{n=1}^N \alpha_n  \beta_{n,k} p_{k,n} \leq \hat{p}_k^{\max},  \:
\sum\nolimits_{n=1}^N \alpha_n \beta_{n,k} f_{n,k}^e \leq f_{k, \max}^e, \\
\label{OP1_9}&   \quad \quad e_n \leq e_n^{\max}, \forall n \in \mathcal{N},
\end{align}
\end{subequations}
where constraints are explained as follows. Constraints \eqref{OP1_3} and \eqref{OP1_4} denote that under the offloading mode, each transmitter is served by one edge server and each edge server can serve at most $N_k^{\max}$ transmitters; \eqref{OP1_8} means the transmit power constraint can not exceed its maximum value $p_n^{\max}$ and the computation frequency of transmitter $t_n$ can not exceed its maximum value $f_{n,\max}^L$, respectively; \eqref{OP1_6} indicates that the overall transmit powers of $A_k$ allocated to its associated transmitters/receivers can not exceed maximum value $\hat{p}_k^{\max}$ and also indicates that the overall computation frequencies of $A_k$ allocated to its associated transmitters' generation tasks is less than the maximum value $f_{k,\max}^e$; \eqref{OP1_9} ensures that the energy consumption of $t_n$ is less than the device energy budget $e_n^{\max}$. {By solving problem \eqref{formulationProblem}, the model selection and communication/computation parameters can be selected and adjusted, which is determined by our proposed algorithm in Section \ref{ALSection}. }
\vspace{-1em}
\section{Proposed SLJ-based matching algorithm}
\label{ALSection}
Recall that problem \eqref{formulationProblem} is an MINLP problem with coupled discrete and continuous variables, making it computationally challenging to obtain the optimal solution within polynomial time. To efficiently solve this problem, we first equivalently decompose it as a two-level problem and then propose an SLJ-based matching algorithm.
Specifically, considering the discrete and continuous characteristics of optimization variables, we split the set of variables of \eqref{formulationProblem} into two subsets, i.e., $\{ \alpha_n, \beta_{n,k} \}$ and $\{ p_n, f_n^L, p_{k,n}, f_{n,k}^e \}$, where the outer level of this problem is to optimize $\{ \alpha_n, \beta_{n,k} \}$ and the inner level of this problem is to optimize $\{ p_n, f_n^L, p_{k,n}, f_{n,k}^e \}$, respectively. 

\subsubsection{Inner level of optimizing $\{ p_n, f_n^L, p_{k,n}, f_{n,k}^e \}$}
Given discrete $\{ \alpha_n, \beta_{n,k} \}$, problem \eqref{formulationProblem} is decomposed as
\begin{equation}
\label{FP2_1}
 \vspace{-2pt}
 \quad \min_{\substack{p_n, f_n^L, p_{k,n}, f_{n,k}^e}}   \max_{\forall n} m_n^{CCQ} \quad  \text{s.t.}  \quad \eqref{OP1_8}\sim\eqref{OP1_9}.
 \vspace{-2pt}
\end{equation}
To address the non-smoothness caused by $\max$ operation in objective function, we introduce an auxiliary variable $\phi$ to transform problem \eqref{FP2_1} as follows
\begin{equation}
\label{OP2_2}
\vspace{-2pt}
 \quad \min_{\substack{p_n, f_n^L, p_{k,n}, f_{n,k}^e}, \phi}  \phi
\quad  s.t. \: \eqref{OP1_8}\sim\eqref{OP1_9}; \: m_n^{CCQ} \leq \phi, \forall n \in \mathcal{N}.
 \vspace{-2pt}
\end{equation}
To tackle complicated terms $\{e_n^{T-R}, e_{n,k}^u\}$ in \eqref{OP1_9}, we reform the transmit power as a function of the communication latency (i.e., $\tau_n^{T-R},\tau_{k,n}^d$, $\tau_{k,n}^d$). If $\alpha_n=0$, based on $\tau_n^{T-R} = \frac{|X_n^\prime|}{R_n^{T-R}}$, we have $p_n \triangleq (2^{|X_n^{\prime}|/(B\tau_n^{T-R})} - 1) \frac{\sigma_n^2}{|h_n|^2 }$. Likewise, if $\alpha_n=1$, we have $p_n \triangleq \sum_{k=1}^K \beta_{n,k}( 2^{|X_n|/(B\tau_{n,k}^u)} -1) \frac{\sigma_{n,k}^2}{|h_{n,k}^u|^2}$, and $p_{k,n} \triangleq (2^{|X_{n,k}^{\prime}|/(B\tau_{k,n}^d)}-1) \frac{\sigma_{k,n}^2}{|h_{k,n}^d|^2}$. Then, by substituting $p_n$ and $p_{k,n}$ in \eqref{OP2_2}, this problem can be transformed into a tractable convex problem that can be solved using existing convex optimization tools, e.g., CVX tools\cite{2024JSAC2}.
\subsubsection{Outer level of optimizing $\{ \alpha_n, \beta_{n,k} \}$} Since the continuous variables of the inner level can be viewed as functions of outer-level discrete variables, problem \eqref{formulationProblem} is equivalent to
\begin{equation}
\label{FP3_1}
 \vspace{-2pt}
 \min_{\substack{\alpha_n, \beta_{n,k}}} \max_{\forall n} m_n^{CCQ} \quad  \text{s.t.}  \quad \eqref{OP1_1}\sim\eqref{OP1_4}, \eqref{OP1_6}, \eqref{OP1_9}
  \vspace{-2pt}
\end{equation}
which is an integer optimization problem. Based on the matching theory \cite{matching}, we transform problem \eqref{FP3_1} as follows.

\emph{Definition 1 (One-to-Many Matching Model):} Two-sided matching $\varphi$ represents the mapping relationship of task offloading and edge server association strategy, satisfying the following conditions: a) $\varphi(n) \in \mathcal{K}\cup \varnothing$,  $|\varphi(n)| = 1, \forall n \in \mathcal{N} $; $\varphi(n) = \varnothing $ if $t_n$ chooses to locally generate the prompt; b) $\varphi(k) \in \mathcal{N} \cup \varnothing$, $|\varphi(k)| \leq N_{k}^{\max}$, $\forall k \in \mathcal{K}$; $\varphi(k) = \varnothing$ if $A_k$ does not serve any T-R pairs; c) $\varphi(n) = k$ if and only if $n \in \varphi(k)$, $\forall n \in \mathcal{N}, \forall k \in \mathcal{K}$. To better describe the matching relationship between transmitters and edge servers, we define the utility function as $U[\varphi] \triangleq \max\limits_{\forall n} m_n^{CCQ}[\alpha_n, \beta_{n,k}, p_n^{\ast}, f_n^{L\ast}, p_{k,n}^{\ast}, f_{n,k}^{e\ast}]$ to measure the performance of different matchings. Then, problem \eqref{FP3_1} can be equivalently transformed as finding the best matching to minimize utility function $U[\varphi]$.

{\color{black}Note that conventional swap-based algorithms \cite{matching} for one-to-many matching problems often overlook the influence of unmatched players, which may lead to local optimality. Herein, we design an SLJ-based matching algorithm with the following swap/leaving/joining operations.}

\emph{Definition 2 (Swap Matching and Swap-Blocking Matching):} For transmitters $t_n$ and $t_{n^{\prime}}$, a \emph{swap matching} is defined as $ \varphi^S_{n,n^{\prime}} \triangleq \{ \varphi \backslash \{ (n, \varphi (n)), ( n^{\prime}, \varphi (n^{\prime})) \} \cup \{ ( n, \varphi (n^{\prime})), (n^{\prime}, \varphi (n))   \} \}$, where transmitters $t_n$ and $t_{n^{\prime}}$ will exchange their currently matched edge servers $A_k$ and $A_{k^{\prime}}$ while keeping other transmitters' matching conditions unchanged. Then, in a feasible matching $\varphi$, $\varphi^S_{n,n^{\prime}}$ can be a \emph{swap-blocking matching} if and only if $U[\varphi^S_{n,n^{\prime}}] < U[\varphi]$, where $(n, n^\prime)$ is a swap-blocking pair.

\emph{Definition 3 (Leaving Matching and Leaving-Blocking Matching):}
Consider that transmitter $t_n$ (corresponding to pair $\{t_n, r_n\}$) is matched with $A_k$, i.e., $\varphi(n) = k$. The \emph{leaving matching} is defined as $\varphi_{n,k}^L \triangleq \{ \varphi \backslash \{ (n, k) \cup (n, \varnothing)\}\} $, meaning that $t_n$ will locally generate the prompt. Accordingly, in a feasible matching $\varphi$, $\varphi_{n,k}^L$ can be a \emph{leaving-blocking matching} if and only if $U[\varphi_{n,k}^L] < U[\varphi]$, where $(n,k)$ is a leaving-blocking pair.

\emph{Definition 4 (Joining Matching and Joining-Blocking Matching):} Consider that transmitter $t_n$ locally generates the prompt, i.e., $\varphi(n) = \varnothing$. If edge server $A_k$ is not fully matched, i.e., $|\varphi(k)| < N_k^{\max}$, the \emph{joining matching} over $t_n$ (corresponding to T-R pair $\{t_n, r_n\}$) and $A_k$ is defined as $\varphi^J_{n, k} \triangleq \{\varphi \backslash (n, \varnothing) \cup (n, k) \}$, indicating that $t_n$ will choose $A_k$ for offloaded prompt generation. In a feasible matching $\varphi$, $\varphi^J_{n, k} $ can be a \emph{joining-blocking matching} if and only if $U[\varphi^J_{n, k} ] < U[\varphi]$, where $(n,k)$ is a joining-blocking pair.

Based on the above definitions, we define that a matching $\varphi$ is \emph{two-sided stable} if there does not exist a swap/leaving/joining-blocking pair. Then, the procedure of the proposed SLJ-based matching algorithm is summarized as follows. \emph{Step a}: A feasible matching $\varphi$ is randomly generated. \emph{Step b}: Each transmitter $t_n$ ($\forall n \in \mathcal{N}$) tries to swap with another transmitter $t_{n^{\prime}}$ ($\forall n^{\prime} \in \{ \mathcal{N} \setminus n \} $) and formulates a new matching $\varphi^S_{n,n^{\prime}}$. Then, update matching $\varphi=\varphi^S_{n,n^{\prime}}$ if $\varphi^S_{n,n^{\prime}}$ is a swap-blocking matching. \emph{Step c}: Each transmitter $t_n$ ($n \in \mathcal{N}$ and $\varphi(n) \neq \varnothing$ (i.e., $\varphi(n) = k$)) tries to leave its current matched server $A_k$ and formulates $\varphi_{n,k}^L$. Then, update $\varphi = \varphi_{n,k}^L$ if $\varphi_{n,k}^L$ is a leaving-blocking matching. \emph{Step d}: Each transmitter $t_n$ ($n \in \mathcal{N}$ and $\varphi = \varnothing$) tries to join in a not fully matched server $k$ ($k \in \mathcal{K}$ and $|\varphi(k)| \leq N_{k}^{\max}$) and formulates $\varphi^J_{n, k}$. Then, update matching $\varphi=\varphi^J_{n, k}$ if $\varphi^J_{n, k}$ is a joining-blocking matching. \emph{Step e}: Repeat \emph{Steps b}$\sim$\emph{d} until there is no swap/leaving/joining-blocking matching and output a two-sided stable matching $\varphi$.

{Note that existing matching algorithms\cite{matching} allow only local swaps between current matched pairs and therefore are highly sensitive to initial matchings. In contrast, the proposed algorithm incorporates leaving and joining operations to expand the search space. Specifically, by enabling transmitters to leave unfavorable edge servers or enabling unmatched transmitters to join new edge servers, the proposed algorithm reduces the dependency on initial matchings, thereby significantly improving the performance.}

{Since the solution of inner-level problem can be obtained using existing convex optimization tools, e.g. CVX toolbox using interior-point method\cite{2024JSAC2}, the convergence and complexity of solving inner-level problem can be guaranteed. Therefore, we focus on the convergence and complexity analysis of the proposed SLJ-based algorithm for the outer-level problem.}

{\emph{Convergence analysis:}  In the proposed algorithm, each transmitter sequentially conducts the swap/leaving/joining operation and generates a new matching. If the newly generated matching reduces the maximal CCQ, the current matching is replaced with this newly generated matching. Therefore, the proposed algorithm can iteratively generate a series of matchings with non-increasing values of maximal CCQ. Moreover, since the number of transmitters and edge servers is limited, indicating that the number of swap/leaving/joining-blocking operations is polynomially countable \cite{4DMatching}. Therefore, it can be concluded that the proposed algorithm can converge to a two-sided stable matching with limited operations/iterations.}

{\emph{Complexity analysis:} Note that \emph{Steps b}$\sim$\emph{d} of the proposed algorithm will stop until there is no swap/leaving/joining-blocking matching. Therefore, the time complexity for solving \eqref{FP3_1} is $\mathcal{O}(L(l_1 N(N-1) + l_2N + l_3NK ))$, where $l_1$, $l_2$, and $l_3$ denote the number of loops corresponding to \emph{Steps b, c} and \emph{d}, respectively, and $L$ denotes the number of loops of \emph{Step e}. Note that $l_1$, $l_2$, $l_3$, and $L$ are generally much smaller than $N$ or $K$, respectively\cite{matching}. According to the analysis in \cite{space_complexity}, the worst-case space complexity of the proposed algorithm is calculated as $\mathcal{O}(NK+N+K)$. }

\vspace{-1em}
\section{Simulation Results}

We evaluate the performance of the proposed M/VLM-based image SemCom framework on Flickr dataset \cite{hodosh2013framing}.
{\color{black}The average data size of the offloaded images (resized to 224$\times$224 and compressed using JPEG with $0.4$ bit per pixel \cite{wallace1991jpeg}) is $\approx 2\times10^4$ bits. The average data size for prompts of length 50 and 75 characters is $400$ and $600$ bits, respectively.} {\color{black}To account for the heterogeneity of the M/VLM models, we report the computation complexity and zero-shot captioning performance (normalized CIDEr) achieved by CapPa M/VLM \cite{tschannen2023image}, with different ViT model architectures and prompt length in TABLE \ref{Heterogenity}, where \textcolor{black}{$Q_{\min}$ and $Q_{\max}$ are empirically set to 55 and 125.8, respectively.} These values are used for the FLOPs, i.e. $F_n$ and $F_k^{\prime}$, and the corresponding performance, i.e. $\hat{Q}_n$ and $\hat{Q}_{n,k}^{\prime}$, respectively. Herein, the device models are randomly selected from the S/16 and M/16, and the edge models from the B/16 and L/14 architectures to account for model heterogeneity.}

The edge servers/transmitters/receivers are randomly located within discs centered at $(250, 250)$/$(0,0)$/$(0,400)$ with radii of $200$/$100$/$100$, respectively. The path loss between nodes $p$ and $q$ is modeled as $1/ \big(1+(d_{p,q}/ d_0) \big)^{\kappa}$, where $\kappa = 2.7$ is the path loss exponent, $d_{p,g}$ denotes the distance between $p$ and $g$, and $d_0=10$ is the reference distance. The bandwidth is set to $B = 2$ MHz and the noise power spectral density is $-174$ dBm/Hz{\cite{tse2005fundamentals}}. The maximum transmit powers of edge servers and transmitters are set to $\hat{p}_k^{\max} = 30$ and $p_n^{\max} = 20$ dBm, respectively. The weight factor is $\omega_n=1$. The computation intensity for conducting one floating point operation is set to $I_n = I_k^{\prime}= 0.01$ cycles$/$FLOP. The maximum computation frequencies of transmitters and edge servers are randomized from uniform distribution over $[f_{\min}, f_{max}]$ and $[\hat{f}_{min}, \hat{f}_{\max}]$ Gcycles/s, where $f_{\min}$/$\hat{f}_{\min}$ and $f_{max}$/$\hat{f}_{\max}$ are lower bound and upper bound of generated frequencies for transmitters/edge servers, respectively. Moreover, $N=K=4$, $N_k^{\max}=3$, $[\hat{f}_{min}, \hat{f}_{\max}]=[11, 14]$, $\kappa_n=10^{-27}$, and $e_n^{\max}=12$ J{\cite{2024JSAC2}}.

\begin{table}[t]
\scriptsize
\centering
\captionsetup{font={footnotesize}, singlelinecheck = off, justification = justified, name={TABLE},labelsep=period}
\caption{\color{black}Computation complexity and zero-shot captioning performance (Normalized CIDEr) for various model sizes/prompt lengths, where models pre-trained on 9 billion image-text pairs.}
\begin{tabular}{|c|c|c|c|c|}
\hline
\begin{tabular}[c]{@{}c@{}}Arch. \\ {(FLOPs)}\end{tabular} & \begin{tabular}[c]{@{}c@{}}S/16\\ { (9.2G)}\end{tabular} & \begin{tabular}[c]{@{}c@{}}M/16\\ {(16.0G)}\end{tabular} & \begin{tabular}[c]{@{}c@{}}B/16\\ { (35.1G)}\end{tabular} & \begin{tabular}[c]{@{}c@{}}L/14\\ { (161.8G)}\end{tabular} \\ \hline
\begin{tabular}[c]{@{}c@{}}Prompt Length: $400$ bits\end{tabular} & 0.03 & 0.10 & 0.20 & 0.31 \\ \hline
\begin{tabular}[c]{@{}c@{}}Prompt Length: $600$ bits\end{tabular} & 0.15 & 0.23 & 0.36 & 0.48 \\ \hline
\end{tabular}
\label{Heterogenity}
\vspace{-1em}
\end{table}

\begin{figure}[t]
		\centering	\includegraphics[width=0.42 \textwidth]{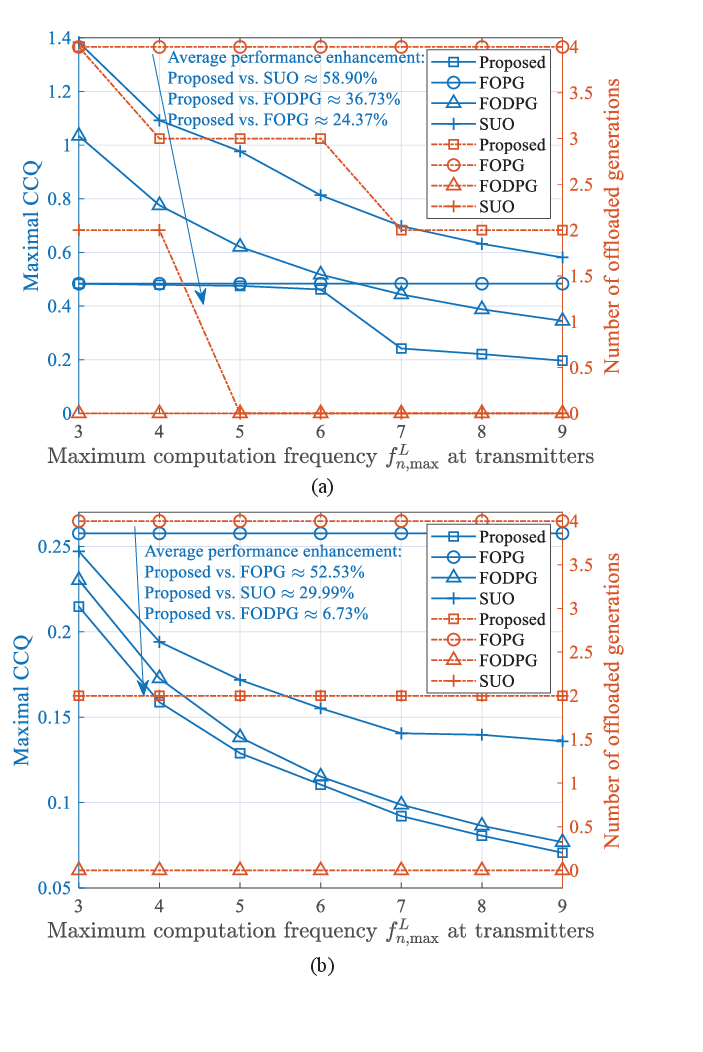}
  \captionsetup{font={footnotesize}, singlelinecheck = off, justification = justified, name={Fig.},labelsep=period}
		\caption{The maximal CCQ and the number of offloaded generations among the proposed framework, FOPG, FODPG, and SUO versus $f_{n,\max}^L$ at transmitters, where
(a) prompt length is $400$ bits, and (b) prompt length is $600$ bits.}
     \vspace{-8mm}
		\label{fig:gausers}
\end{figure}

We compare our proposed framework with the following three benchmarks:  1) \emph{\textcolor{black}{Fully offloaded prompt generation (FOPG),}} where each transmitter offloads its image to a pre-selected edge server, then the edge server generates the prompt and transmits it to the paired receiver. 2) \emph{\textcolor{black}{Full on-device prompt generation (FODPG),}} where each transmitter locally generates the prompt using its lightweight model, and directly transmits the prompt to the paired receiver. 3) \emph{\textcolor{black}{Semantic un-aware offloading (SUO)}}, where the maximal latency among T-R pairs is minimized by optimizing edge server selection, communication, and computation resource allocation.

Fig. \ref{fig:gausers} compares the maximal CCQ and the number of offloaded generations of the proposed framework with the benchmarks under varying $f_{n,\max}^L$. It can be seen from this figure that our proposed framework outperforms other benchmarks in terms of the maximal CCQ, \textcolor{black}{on average reducing the maximal CCQ by 38.45\%, 21.73\%, and 44.45\% in comparison with FOPG, FODPG, and SUO schemes over the range of local computation frequency $[3, 9]$ Gcycles/s and prompt length $X^\prime_{n}/X^\prime_{n,k} \in \{400, 600\}$ bits\footnote{{The overall average performance enhancement of the proposed framework compared to FOPG, FODPG, and SUO is calculated as follows: $(24.37\%+52.53\%)/2=38.45\%$, $(36.73\%+6.73\%)/2 =21.73\%$, and $(58.90\%+29.99\%)/2 = 44.445\% \approx44.45\%$, respectively. }}}. It indicates that our proposed algorithm can better decide the prompt generation offloading strategy and select edge servers for generating the prompts. Moreover, the maximal CCQs of the proposed framework, FODPG, and SUO benchmarks decrease with maximum computation frequency $f_{n, \max}^L$ at transmitters. This is because when
the edge servers have sufficient computing resources compared
to the local devices, they can offer lower prompt generation latency and higher quality.
\textcolor{black}{However, when on-device computation resources are sufficient, some transmitters may choose to locally generate the prompt to reduce generation latency in our proposed framework. }
{As shown in Figs. \ref{fig:gausers}(a) and \ref{fig:gausers}(b), the number of offloaded generations in our proposed framework decreases with the prompt length when $f^L_{n,\max} \leq 7$ Gcycles/s. This is because for sufficiently large prompt length, the semantic quality for on-device generation becomes sufficiently good and thereby it is not worth to tolerate the additional communication/computation latency for the small improvement in CIDEr by offloading.}

 \begin{figure}[t]
    \vspace{-1mm}
		\centering	\includegraphics[width=0.47\textwidth]{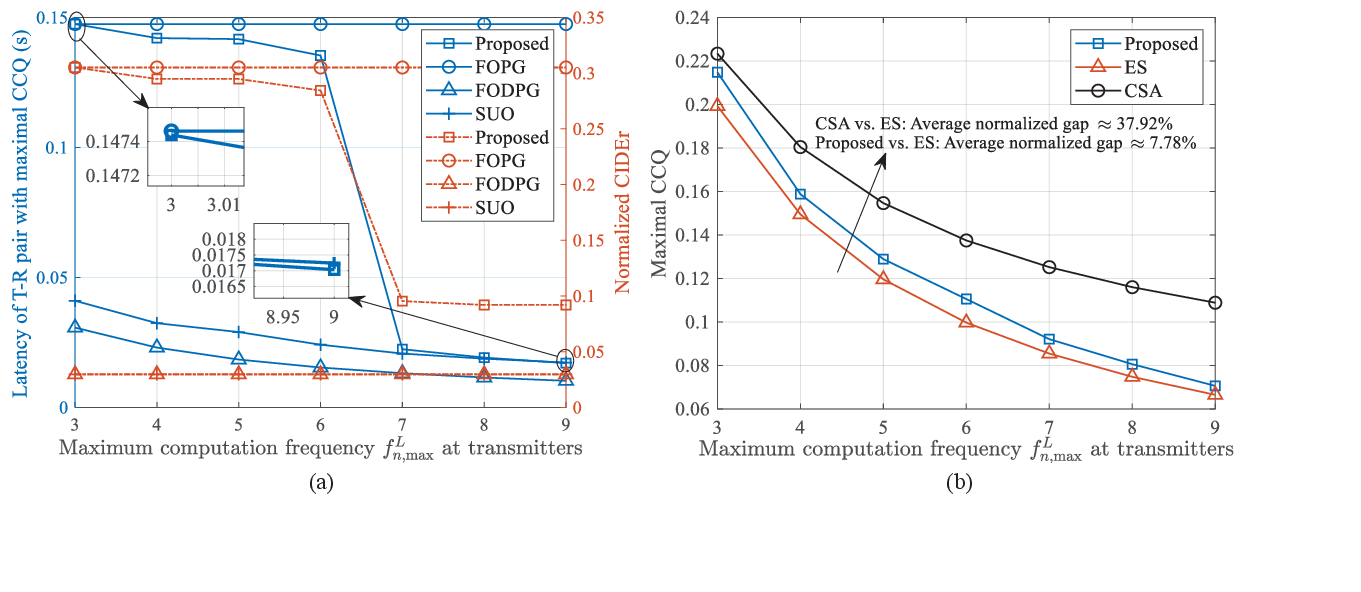}
  \captionsetup{font={footnotesize}, singlelinecheck = off, justification = justified, name={Fig.},labelsep=period}
		\caption{{(a) The normalized CIDEr/latency of T-R pair with maximal CCQ of the proposed framework and other schemes, where prompt length is $ 400$ bits. (b) The maximal CCQ of the proposed framework, ES, and CSA versus $f_{n,\max}^L$ at transmitters, where prompt length is $600$ bits.}}
    \vspace{-8mm}
		\label{fairness}
    \end{figure}

{Fig. \ref{fairness}(a) depicts the latency\footnote{{To be consistent with the parameter settings in this paper, the following examples for practical chips deployed on edges and devices are provided. When the edge server adopts GeForce MX250, Intel Iris Plus Graphics G7 (64EU) or Radeon 740M chip (1.08$\sim$1.43 TFLOPs/s), the computation latency for large models (e.g., 35.1/161.8 GFLOPs) approximately ranges from 0.0245 s to 0.1498 s. When the local device employs Intel UHD Graphics 730, Intel UHD Graphics Xe G4 (48 EU) or GeForce MX230 (0.35$\sim$1.08 TFLOPs/s), the computation latency for local light-weight models (e.g., 9.2/16.0 GFLOPs) ranges from 0.0085 s to 0.0457 s. These values are comparable to those shown in Fig. 3(a), indicating that the presented simulation results of latency can provide a reference for practical real-time deployment of models.}} and the normalized CIDEr of the T-R pair with maximal CCQ of the proposed framework and other schemes versus $f^{L}_{n,\max}$. It can be seen that the latency of proposed framework is lower than that of FOPG, and the normalized CIDEr of proposed framework is higher than those of FODPG and SUO schemes. This means that a better semantic performance is achieved at the cost of longer communication/computation latency. Moreover, this observation indicates that the proposed framework is more effective in balancing latency and semantic quality than other schemes.} {Fig. \ref{fairness}(b) compares the maximal CCQ of the proposed framework, the convention swap-based algorithm (CSA)\cite{matching}, and the exhaustive search (ES), where the normalized gap is defined as the ratio of the maximal CCQ achieved by proposed/CSA to that achieved by ES. It can be observed that the normalized gap between proposed framework and ES is lower than that between CSA and ES, indicating that our proposed algorithm can effectively enhance the performance compared with CSA.}

\vspace{-1em}
\section{Conclusions}
In this paper, we proposed an edge-device collaborative Gen SemCom framework via textual prompts leveraging pre-trained M/VLMs. We optimized prompt generation offloading, and the communication and computation resource allocation to reduce the latency while increasing the semantic quality in a multiple transmitter/receiver setup. We proposed an SLJ-based matching algorithm for solving the formulated MINLP problem. Simulation results demonstrated the superiority of the proposed framework over the semantic-unaware benchmarks.

\vspace{-1em}
\bibliographystyle{IEEEtran}

\begin{thebibliography}{10}
\providecommand{\url}[1]{#1}
\csname url@samestyle\endcsname
\providecommand{\newblock}{\relax}
\providecommand{\bibinfo}[2]{#2}
\providecommand{\BIBentrySTDinterwordspacing}{\spaceskip=0pt\relax}
\providecommand{\BIBentryALTinterwordstretchfactor}{4}
\providecommand{\BIBentryALTinterwordspacing}{\spaceskip=\fontdimen2\font plus
\BIBentryALTinterwordstretchfactor\fontdimen3\font minus \fontdimen4\font\relax}
\providecommand{\BIBforeignlanguage}[2]{{%
\expandafter\ifx\csname l@#1\endcsname\relax
\typeout{** WARNING: IEEEtran.bst: No hyphenation pattern has been}%
\typeout{** loaded for the language `#1'. Using the pattern for}%
\typeout{** the default language instead.}%
\else
\language=\csname l@#1\endcsname
\fi
#2}}
\providecommand{\BIBdecl}{\relax}
\BIBdecl

\bibitem{TelecomGen}
L.~Bariah \emph{et~al.}, ``Large generative {AI} models for telecom: The next big thing?'' \emph{IEEE Commun. Mag.}, vol. 62, no. 11, pp. 84--90, Nov. 2024.

\bibitem{NetGPT}
Y.~Chen \emph{et~al.}, ``{NetGPT}: An {AI}-native network architecture for provisioning beyond personalized generative services,'' \emph{IEEE Netw.}, vol. 38, no. 6, pp. 404--413, Nov. 2024.

\bibitem{LLM}
H.~Zhou \emph{et~al.}, ``{Large Language Model (LLM)} for telecommunications: A comprehensive survey on principles, key techniques, and opportunities,'' \emph{{IEEE} Commun. Surveys Tuts.}, DOI: 10.1109/COMST.2024.3465447.

\bibitem{Liang2024generative}
C.~Liang \emph{et~al.}, ``Generative {AI}-driven semantic communication networks: Architecture, technologies and applications,'' \emph{IEEE Trans. Cogn. Commun. Net.}, vol.~11, no.~1, pp. 27--47, Feb. 2025.

\bibitem{Wanting2024generative}
W.~Yang \emph{et~al.}, ``Rethinking generative semantic communication for multi-user systems with multi-modal {LLM},'' \emph{arXiv:2408.08765v1}, 2024.

\bibitem{10343094}
C.~Liang \emph{et~al.}, ``Selection-based image generation for semantic communication systems,'' \emph{{IEEE} Commun. Lett.}, vol.~28, no.~1, pp. 34--38, Jan. 2024.

\bibitem{qiao2024latency}
L.~Qiao \emph{et~al.}, ``Latency-aware generative semantic communications with pre-trained diffusion models,'' \emph{IEEE Wireless Commun. Lett.}, vol. 13, no. 10, pp. 2652--2656, Oct. 2024.

\bibitem{Zhijin}
H.~Xie \emph{et~al.}, ``Towards intelligent communications: Large model empowered semantic communications,'' \emph{IEEE Commun. Mag.}, vol. 63, no. 1, pp. 69--75, Jan. 2025.

\bibitem{cicchetti2024language}
G.~Cicchetti \emph{et~al.}, ``Language-oriented semantic latent representation for image transmission,'' in \emph{Proc. 2024 IEEE 34th Int. Workshop Mach. Learn. Signal Process. (MLSP)}, London, UK, 2024, pp. 1--6.



\bibitem{2024TVT}
{L.~Zhong \emph{et~al.}, ``Distributed optimization of multi-role {UAV} functionality switching and trajectory for security task offloading in {UAV}-assisted {MEC},'' \emph{{IEEE} Trans. Veh. Technol.}, vol.~73, no.~12, pp. 19432--19447, Dec. 2024.}

\bibitem{2024JSAC1}
{Y.~Peng \emph{et~al.}, ``Stochastic long-term energy optimization in digital twin-assisted heterogeneous edge networks,'' \emph{{IEEE} J. Sel. Areas Commun.}, vol.~42, no.~11, pp. 3157--3171, Nov. 2024.}

\bibitem{2024JSAC2}
{X.~Yu \emph{et~al.}, ``Joint resource allocations for energy consumption optimization in {HAPS}-aided {MEC-NOMA} systems,'' \emph{{IEEE} J. Sel. Areas Commun.}, vol.~42, no.~12, pp. 3632--3646, Dec. 2024.}

\bibitem{SemCom1}
D.~G\"{u}nd\"{u}z \emph{et~al.}, ``Beyond transmitting bits: Context, semantics, and task-oriented communications,'' \emph{{IEEE} J. Sel. Areas Commun.}, vol.~41, no.~1, pp. 5--41, Jan. 2023.

\bibitem{Semcom2}
W.~Yang \emph{et~al.}, ``Semantic communications for future internet: Fundamentals, applications, and challenges,'' \emph{{IEEE} Commun. Surveys Tuts.}, vol.~25, no.~1, pp. 213--250, 1st Quat., 2023.

\bibitem{FSL1}
T.~B. Brown \emph{et~al.}, ``Language models are few-shot learners,'' in \emph{Proc. Adv. Neural Inf. Process. Syst. (NeurIPS)}, 2020.


\bibitem{matching}
{W. He \emph{et~al.}, ``Joint user association, resource allocation, and beamforming in RIS-assisted multi-server MEC systems,'' \emph{IEEE Trans. Wireless Commun.}, vol. 23, no. 4, pp. 2917--2932, Apr. 2024.}



\bibitem{integ}
{L. Xia \emph{et~al.}, ``Generative AI for semantic communication: Architecture, challenges, and outlook,'' \emph{IEEE Wireless Commun.}, vol. 32, no. 1, pp. 132--140, Feb. 2025.}

\bibitem{AutoRT}
{M.~Ahn \emph{et~al.}, ``{AutoRT}: Embodied foundation models for large scale orchestration of robotic agents,'' \emph{arXiv:2401.12963}, 2024.}


\bibitem{XR}
{W.~Yang \emph{et~al.}, ``Streamlined transmission: A semantic-aware {XR} deployment framework enhanced by generative {AI},'' \emph{IEEE Netw.}, vol. 38, no. 6, pp. 29--38, Nov. 2024. }

\bibitem{IoS}
{N.~Sehad \emph{et~al.}, ``Generative {AI} for immersive communication: The next frontier in internet-of-senses through {6G},'' \emph{IEEE Commun. Mag.}, vol. 63, no. 2, pp. 31--43, Feb. 2025.}



\bibitem{yergeau2003utf}
{F.~Yergeau, ``{UTF}-8, a transformation format of {ISO} 10646,'' Tech. Rep., 2003.}

\bibitem{wallace1991jpeg}
{G.~K. Wallace, ``The {JPEG} still picture compression standard,'' \emph{Communications of the ACM}, vol.~34, no.~4, pp. 30--44, 1991.}




\bibitem{4DMatching}
M.~Ren \emph{et~al.}, ``Energy-delay tradeoff in helper-assisted noma-mec systems: A four-sided matching algorithm,'' \emph{{IEEE} Trans. Commun.}, vol.~72, no.~5, pp. 2835--2850, May 2024.

\bibitem{space_complexity}
A.~R. Usmani, ``A novel time and space complexity efficient variant of counting-sort algorithm,'' in \emph{Proc. 2019 Int. Conf. Innovative Computing (ICIC)}, 2019, pp. 1--6.

\bibitem{hodosh2013framing}
M.~Hodosh, P.~Young, and J.~Hockenmaier, ``Framing image description as a ranking task: Data, models and evaluation metrics,'' \emph{J. Artif. Intell. Res.}, vol.~47, pp. 853--899, Aug. 2013.



\bibitem{tschannen2023image}
{M.~Tschannen \emph{et~al.}, ``Image captioners are scalable vision learners too,'' in \emph{Neural Information Processing Systems (NeurIPS)}, 2023.}


\bibitem{tse2005fundamentals}
{D.~Tse and P.~Viswanath, \emph{Fundamentals of wireless communication}.\hskip 1em plus 0.5em minus 0.4em\relax Cambridge university press, 2005.}


\end{thebibliography}

\end{document}